# Assessment of a GOES Microburst Product for Two Early Cold Season Convective Storms

Kenneth L. Pryor
Center for Satellite Applications and Research (NOAA/NESDIS)
Camp Springs, MD

## 1. Introduction

On 17 November and 1 December 2010, lines of convective storms developed ahead of a strong cold front and tracked through the Mid-Atlantic region. In both cases, squall lines produced strong downburst winds as they tracked over the Tidal Potomac River and Chesapeake Bay regions. NOAA and WeatherFlow marine and buoy meteorological observations as well as DCNet observations were instrumental in verifying downburst occurrence and magnitude with both squall lines. This paper presents an assessment of the Geostationary Operational Environmental Satellite (GOES) imager channel 3 - 4 brightness temperature difference (BTD) product for these two early cold-season convective storm and downburst events. It has been found recently that the BTD between GOES infrared channel 3 (water vapor at 6.5μm) and channel 4 (thermal infrared at 11μm) can highlight regions where severe outflow wind generation (i.e. downbursts, microbursts) is likely due to the channeling of dry mid-tropospheric air into the precipitation core of a deep, moist convective storm (Pryor 2010).

Cold season convective storm events are the result of a somewhat different thermodynamic environment than that associated with the warm season: marginal static instability and strong low- and mid-level wind shear are typically the dominant forcing factors for strong convective winds associated with cold season convective storms. The 17 November and 1 December cases exemplify dynamically-forced convective systems with well-developed rear-inflow jets.

Weisman (1992) describes the physical process of the development of a rear-inflow jet (RIJ) in long-lived convective storms. As shown in Figure 1 and discussed in the following case studies, strong low-to-mid-level wind shear will be very important in generating the storm-scale vertical circulation that results in the development of an RIJ. Subsequently, the RIJ is primarily responsible for the channeling of dry air into the precipitation cores of convective storms and provides the downdraft energy necessary to result in strong downburst winds associated with the squall lines.

## 2. Methodology

The objective of this validation effort was to assess the performance of the GOES imager channel 3-4 BTD in the indication of downburst activity during cold-season convective storm events by applying pattern recognition techniques. The image data consisted of derived brightness temperatures from infrared bands 3 and 4, obtained from the Comprehensive Large Array-data Stewardship System (CLASS, http://www.class.ncdc.noaa.gov/). Microburst algorithm output was visualized by McIDAS-V software (Available online at http://www.ssec.wisc.edu/mcidas/software/v/). A contrast stretch and color enhancement were applied to the output images to highlight patterns of interest including overshooting tops and dry-air notches. The dry-air notch identified in the BTD image is similar in concept to the rear-inflow notch (RIN) as documented in Przybylinski (1995). Visualizing algorithm output in

McIDAS-V allowed for cursor interrogation of output brightness temperature and more precise observing of BTD values associated with downburst events.

For each downburst event, product images were compared to radar reflectivity imagery and surface observations of convective wind gusts as provided by WeatherFlow, the National Data Buoy Center (NDBC), and DCNet. An expanded suite of high-quality boundary layer measurements is available with DCNet observations, thus providing a deeper analysis of the physical process of downburst generation. DCNet stations record physical quantities such as maximum wind speed for zonal, meridional, and vertical (U, V, W) components, wind fluctuation variance for zonal, meridional, and vertical ($u'^2, v'^2, u'^2$) components, and turbulent vertical momentum fluxes ($u'w', v'w'$). The observation of these parameters allow for a more effective description of the storm-scale wind field associated with a downburst. 15-minute maximum wind speeds for each downburst event were derived from component wind measurements by employing a mathematical formula that sums the square of each horizontal wind component. In addition, Next Generation Radar (NEXRAD) base reflectivity imagery acquired from National Climatic Data Center (NCDC) was utilized to verify that observed wind gusts were associated with downbursts and not associated with other types of convective wind phenomena (i.e. gust fronts). Another application of the NEXRAD imagery was to infer physical properties of downburst-producing convective storms. Particular radar reflectivity signatures, such as the bow echo and rear-inflow notch (RIN)(Przybylinski 1995), were effective indicators of the occurrence of downbursts.

## 3. Case Studies

### 3.1 17 November 2010 Squall Line Downbursts

During the late evening of 16 November and early morning of 17 November 2010, lines of strong convective storms developed and tracked eastward over Pennsylvania, Maryland and Virginia. The line of convective storms that developed along a cold front boundary produced strong downburst winds over the Tidal Potomac and Chesapeake Bay regions.

The Tidal Potomac Region was affected by the squall line between 0545 and 0630 UTC, as evidenced by high winds recorded by NOAA data buoys and WeatherFlow observing stations. During this time, several dry-air channels became apparent in BTD imagery on the western flank of the convective line. The dry-air channels, pointing into the rear flank of the convective storm line, fostered downburst generation by injecting drier (unsaturated) air into the precipitation core of the convective storm. The resulting evaporational cooling and generation of negative buoyancy accelerated the convective storm downdraft toward the surface, producing strong winds on impact.

Figure 2 shows the squall line moving into the Washington, DC metropolitan area with the apex of a bow echo moving into southern Maryland. A severe wind gust of 50 knots was recorded by the Upper Potomac River buoy (white cross in Figure 1) between 0540 and 0550 UTC. Note that a dry-air notch was apparent on the western flank of the squall line pointing east-northeastward toward the Upper Potomac buoy. At this time, the dry-air notch was in phase with rear-inflow notch (RIN) as evident in NEXRAD imagery. The 0000 UTC 17 November 2010 radiosonde observation (RAOB) from Dulles Airport, Virginia in Figure 3 shows a dry-air layer between the 500 and 600-mb levels, located immediately above a layer of strong low-level wind speed shear. The vertical circulation resulting from the strong low-level wind speed shear as illustrated in Figure 1 as well as the strong convective steering-layer flow between the 900 and 500-mb levels most likely enhanced RIJ strength associated with this squall line.  The well-developed dry-air notches as apparent in GOES imagery and corresponding RINs as apparent in radar imagery were realizations of these enhanced RIJs.

By 0602 UTC two dry-air notches were pointing toward the location of downburst occurrence at Potomac Light 33 station where a wind gust of 43 knots was recorded. About 20 minutes later, near 0625 UTC, a stronger downburst wind gust of 45 knots was recorded by Cobb Point station. At this time, as displayed in Figure 4, a dry-air notch had come in phase with a bow echo moving over the lower Potomac River. It is evident that the injection of mid-tropospheric dry air by a RIJ into the heavy precipitation core within the convective storm line was providing a large amount of downdraft energy to realize strong downburst winds at the surface.  Near this time, as a bow echo was tracking over the Chesapeake Bay, a downburst wind gust of 46 knots was recorded at Thomas Point Lighthouse Coastal-Marine Automated Network (C-MAN) station. Embedded within the bow echo was a spearhead echo that was associated with the particularly intense downburst. Fujita and Byers (1977) closely related the spearhead echo signature to rapidly-moving convective storm cells that generate intense, localized downdrafts. The Greenbury Point WeatherFlow station recorded a less intense wind gust of 37 knots as the apex of the bow echo tracked overhead. Figure 5 shows the passage of the bow echo over the Annapolis Harbor and the associated dry-channel (white line) that was pointing toward the upper Chesapeake Bay in phase with a RIN as apparent in radar imagery. Again, the channeling of dry air into the rear flank of the bow echo was a major forcing factor for downburst winds that were observed in the vicinity of the Annapolis Harbor.  The difference in wind direction between the wind gust observations at Greenbury Point and Thomas Point (southerly) and the prevailing wind direction (southwesterly) in the dry-air layer perhaps reflects the divergent nature of the convective outflow characteristic of a downburst.  Downburst wind gust observations for this event are summarized in Table 1.

Interestingly, strong downbursts, as recorded by DCNet stations at the Naval Research Laboratory and in Silver Spring, Maryland (just north of Washington, DC), also occurred behind the leading convective storm line in association with embedded heavy shower activity near the rear flank of the squall line. As shown in the GOES BTD product image in Figure 6, dry-air notches were pointing directly toward the Naval Research Laboratory and Silver Spring observing stations where downburst wind gusts of 40 knots were recorded between 0600 and 0615 UTC.  The microscale structure of the downbursts was apparent in selected quantities measured by the DCNet stations.  Fujita (1981) identified three components of a microburst observed during the NIMROD project that are displayed in Figure 7:  (1) the "burst front" that marks the leading edge of the downburst wind field; (2) the "burst flow" that incorporates the strongest horizontal winds associated with the microburst; and (3) the "downflow" associated with the intense, localized downdraft.  The DCNet stations effectively observed both the onset (burst front) and maximum horizontal winds (burst flow) associated with the squall line downbursts.  The onset (passage of the burst front) of downburst winds was marked in Figure 8 by a sharp increase in wind fluctuation (gustiness) at 0600 UTC as well as by a sudden increase in both horizontal (V) and vertical (W) components of the wind.  Also shown in Figure 8, the maximum in the horizontal component of the downburst winds (burst flow) was observed between 0600 and 0615 UTC.  During this time, the maxima in the vertical wind component and down-gradient (positive upward) horizontal momentum transport most likely resulted from cyclonic vorticity induced by strong vertical wind shear as exemplified in Figure 7.

## 3.2 1 December 2010 Squall Line Downbursts

The windstorm that evolved along and ahead of a cold front over the Mid-Atlantic region during the morning of 1 December 2010 culminated in the development of a squall line that produced downburst winds over the Tidal Potomac and Chesapeake Bay region. Between 1400 and 1430 UTC, a line of convective storms developed along the cold front over northern Virginia and then tracked east and northeastward into central and southern Maryland between 1430 and 1500 UTC. Between 1430 and 1445 UTC, several strong downbursts occurred over the Tidal Potomac River and were recorded by WeatherFlow observing stations. Similar to the 17 November downburst event, the channeling of mid-tropospheric dry air into the core of convective storms by a RIJ likely provided significant downdraft energy that resulted in high surface winds.

Figure 9 shows a GOES-13 BTD microburst risk image over Maryland and Virginia near the time of downburst occurrence over the Tidal Potomac River. At 1430 UTC, a wind gust of 41 knots was recorded by Potomac Light 33 WeatherFlow station associated with a squall line downburst. Five minutes later, at 1435 UTC, a stronger downburst wind gust of 43 knots was recorded by Cobb Point station. Note that the dry-air notch identified in Figure 9 on the southwestern flank of the squall line was pointing toward Cobb Point, signifying that the injection of dry air into the convective storm was resulting in evaporational cooling and the generation of negative buoyancy that would result in the downburst winds over the Tidal Potomac.

Figure 10 displays favorable conditions for strong convective winds over the Tidal Potomac and Chesapeake Bay region. The sounding profile identifies a dry-air layer near the 800-mb level with winds ranging from 45 to 55 knots, very close to the wind gust speeds observed on the Tidal Potomac and Chesapeake Bay. Although this dry air layer was found at a lower level than indicated in the previous case, it appeared immediately above a low-level wind shear layer. The sounding does provide evidence that in addition to precipitation loading and evaporational cooling, the downward transfer of horizontal momentum from the dry-air layer to the surface by heavy rainfall was also an important forcing factor in the strong downburst winds. As the squall line continued to track eastward, strong downburst winds were observed over the Chesapeake Bay between 1500 and 1530 UTC. At 1517 UTC, Thomas Point Lighthouse Coastal-Marine Automated Network (C-MAN) station recorded a wind gust of 50 knots with the passage of the squall line.

By this time, the dry-air notch on the southwestern flank of the line had become more pronounced in both satellite and radar imagery with a southwest to northeast orientation toward Thomas Point Light (white cross in Figure 11). The peak wind gust recorded at the lighthouse was from a southwesterly direction with a local maximum in radar reflectivity (> 40 dBZ) overhead. Nearby WeatherFlow observing stations, Tolly Point and Greenbury Point, recorded wind gusts of 44 knots and 46 knots, respectively, at 1520 UTC. Table 2 summarizes convective high wind gust observations with this event. The deviation in wind gust direction at Greenbury Point, about five miles north of Thomas Point, from the wind direction in the dry-air layer (southwesterly) again may signify the divergent nature of the convective storm outflow associated with a downburst. The concurrent peak wind gust and passage of the heaviest rainfall core over the observing stations suggest that higher momentum from the mid-tropospheric dry air layer was being transported to the surface by heavy precipitation within the squall line. The physical process in downburst generation described with this event exemplifies the dynamic environment that fostered RIJ development that is typical for cold-season severe convective storms.

In a similar manner to the 17 November event, DCNet stations observed characteristics of a downburst. A downburst wind gust of 41 knots was recorded by the Naval Research Laboratory (NRL) DCNet station at 1445 UTC as the southern end of the squall line was passing overhead. Figure 12 clearly shows the presence of a rear-inflow jet (marked by the white line), pointing toward Washington, DC, that resulted in strong downdraft generation in accordance with the conceptual model presented in Figure 1. As apparent in Figure 13, the occurrence of the downburst over NRL was most effectively indicated by the histograms of the meridional and vertical wind components, wind fluctuation variance ($v'^2$), and horizontal (u) momentum transport where sharp peaks in all of the quantities were observed concurrently with the downburst. This observation underscores the turbulent, gusty nature of downburst winds, identified by Fujita (1981), that features a prominent upward vertical component.

**4. Discussion**

The dry-air notch identified in both cases presented above most likely represents drier (lower relative humidity) air that was channeled into the rear of convective storms and subsequently provided the energy for intense downdrafts and resulting downburst winds. Dry air channeling was likely enhanced by the formation of a RIJ that resulted from strong low-level wind shear, as exemplified in Figure 1. In both cases, comparison of BTD product imagery to corresponding radar imagery revealed a correlation between the dry-air notch and the RIN. Entrainment of drier mid-tropospheric air into the precipitation core of the convective storm typically results in evaporation of precipitation, the subsequent cooling and generation of negative buoyancy (sinking air), and resultant acceleration of a downdraft. When the intense localized downdraft reaches the surface, air flows outward as a downburst. Inspection of proximity RAOBs revealed that downward transport of horizontal momentum from the dry-air layer to the surface was an important factor in downburst wind gust magnitude. This was evidenced by the close correspondence between wind velocity in the dry-air layer and the magnitude of measured downburst wind gusts. These convective high-wind events were typical for the cold season with a large amount of shear and storm precipitation content that provided the forcing for intense downburst winds.

**5. References**

Fujita, T.T., 1981: Tornadoes and downbursts in the context of generalized planetary scales. *J. Atmos. Sci.*, **38**, 1523-1534.

Fujita, T.T., and H.R. Byers, 1977: Spearhead echo and downbursts in the crash of an airliner. *Mon. Wea. Rev.*, **105**, 1292-146.

Pryor, K. L., 2010: Microburst applications of brightness temperature difference between GOES Imager channels 3 and 4. arXiv:1004.3506v1 [physics.ao-ph]

Przybylinski, R.W., 1995: The bow echo. Observations, numerical simulations, and severe weather detection methods. *Wea. Forecasting*, **10**, 203-218.

Weisman, M.L., 1992: Role of Convectively Generated Rear-Inflow Jets in the Evolution of Long-Lived Mesoconvective Systems. *J. Atmos. Sci.*, **49**, 1826–1847.

**Acknowledgements**

The author thanks Jay Titlow, WeatherFlow, Inc., and the NOAA/Air Resources Laboratory [Atmospheric Turbulence and Diffusion Division (ATDD)](#) for the surface weather observation data used in this research effort.

| Location | Network | Time (UTC) | Wind Gust Dir.(º) /Speed (kt) |
|---|---|---|---|
| Upper Potomac Buoy | NOAA/CBIBS | 0550 | 207/50 |
| Potomac Light 33 | WeatherFlow | 0600 | 247/43 |
| Cobb Point | WeatherFlow | 0625 | 250/45 |
| Thomas Point Light | NOAA/C-MAN | 0628 | 170/46 |
| Greenbury Point | WeatherFlow | 0630 | 176/37 |
| Sandy Point | WeatherFlow | 0640 | 212/36 |

Table 1. Selected downburst wind gust observations on 17 November 2010.

| Location | Network | Time (UTC) | Wind Gust Dir.(º) /Speed (kt) |
|---|---|---|---|
| Potomac Light 33 | WeatherFlow | 1430 | 226/41 |
| Cobb Point | WeatherFlow | 1435 | 247/43 |
| Thomas Point Light | NOAA/C-MAN | 1517 | 210/50 |
| Greenbury Point | WeatherFlow | 1520 | 178/46 |
| Tolly Point | WeatherFlow | 1520 | 244/44 |

Table 2. Selected downburst wind gust observations on 1 December 2010.

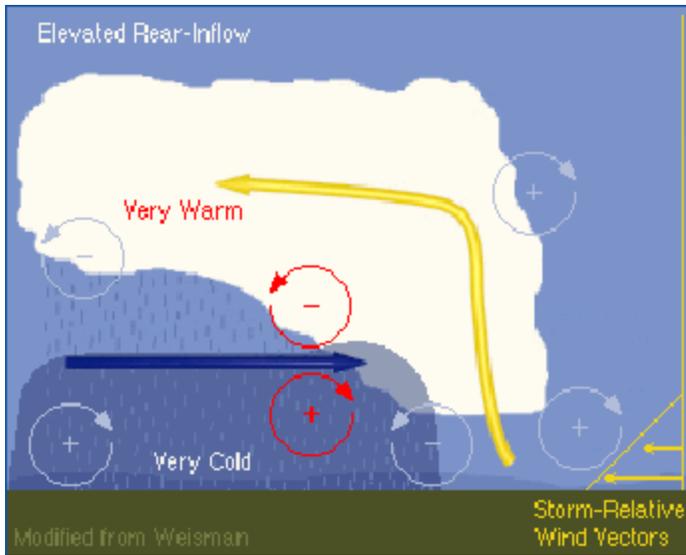

Figure 1. Conceptual model of a non-descending rear-inflow jet. Adapted from Weisman (1992).

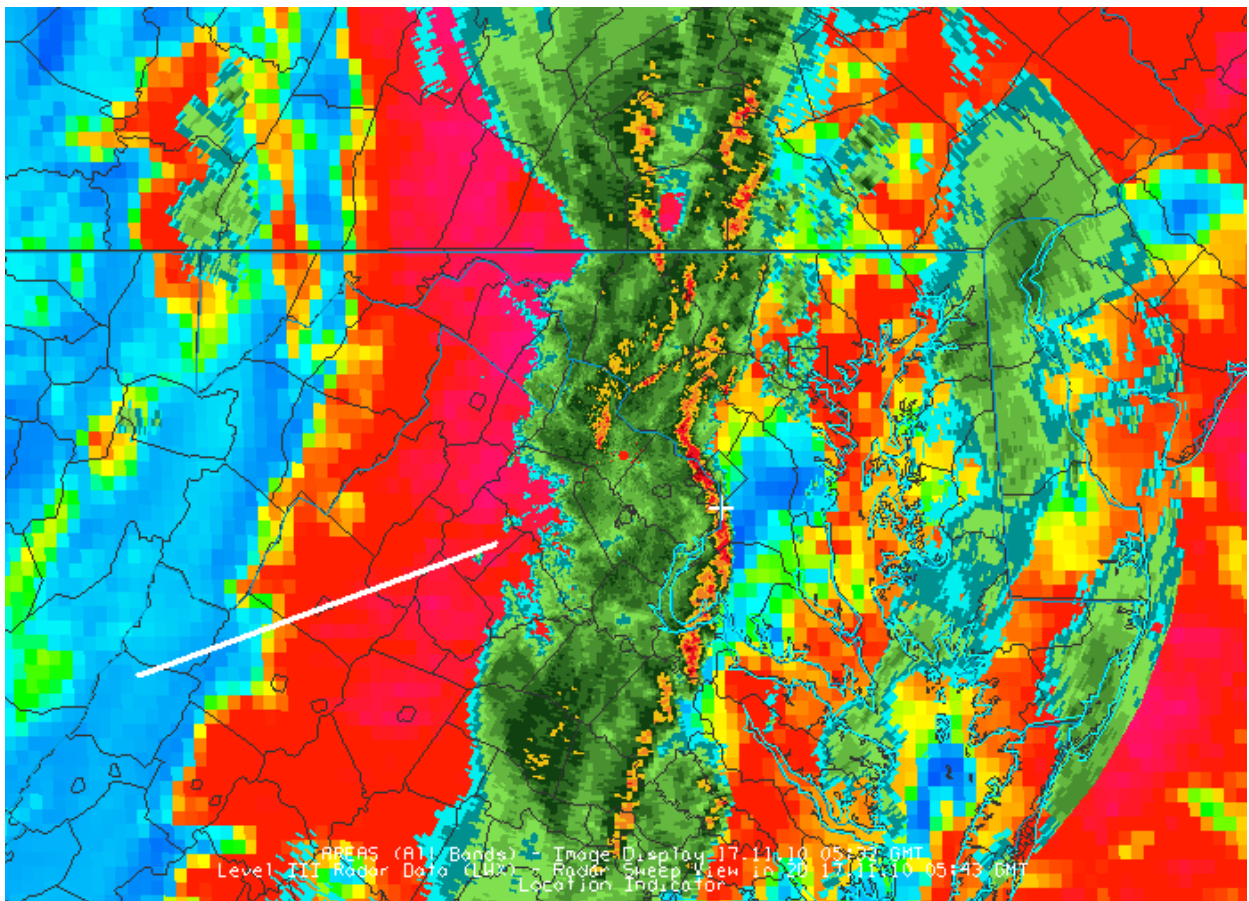

Figure 2. GOES-13 WV-IR BTD image at 0532 UTC 17 November 2010 with overlying radar reflectivity from Sterling, Virginia NEXRAD. White line represents the dry-air notch.

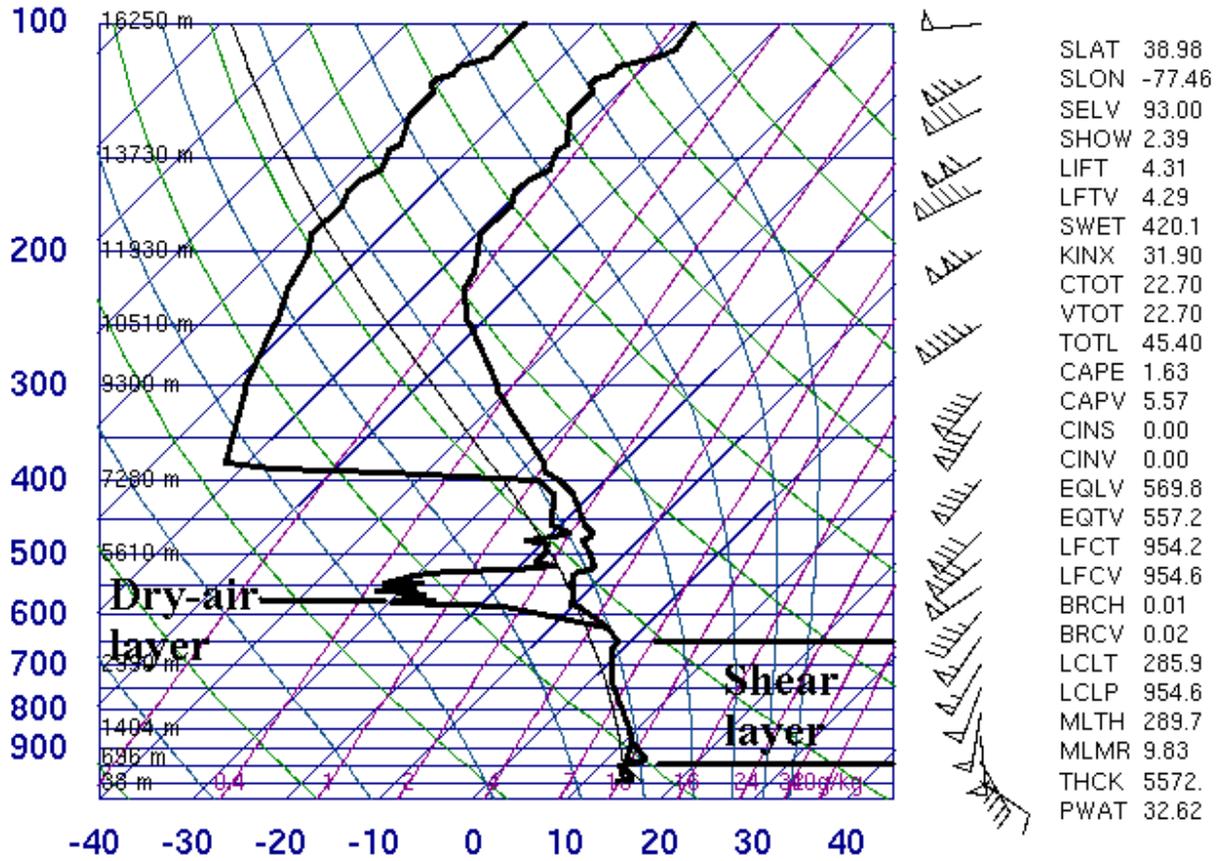
Figure 3. Radiosonde observation (RAOB) from Dulles Airport, VA at 0000 UTC 17 November 2010.

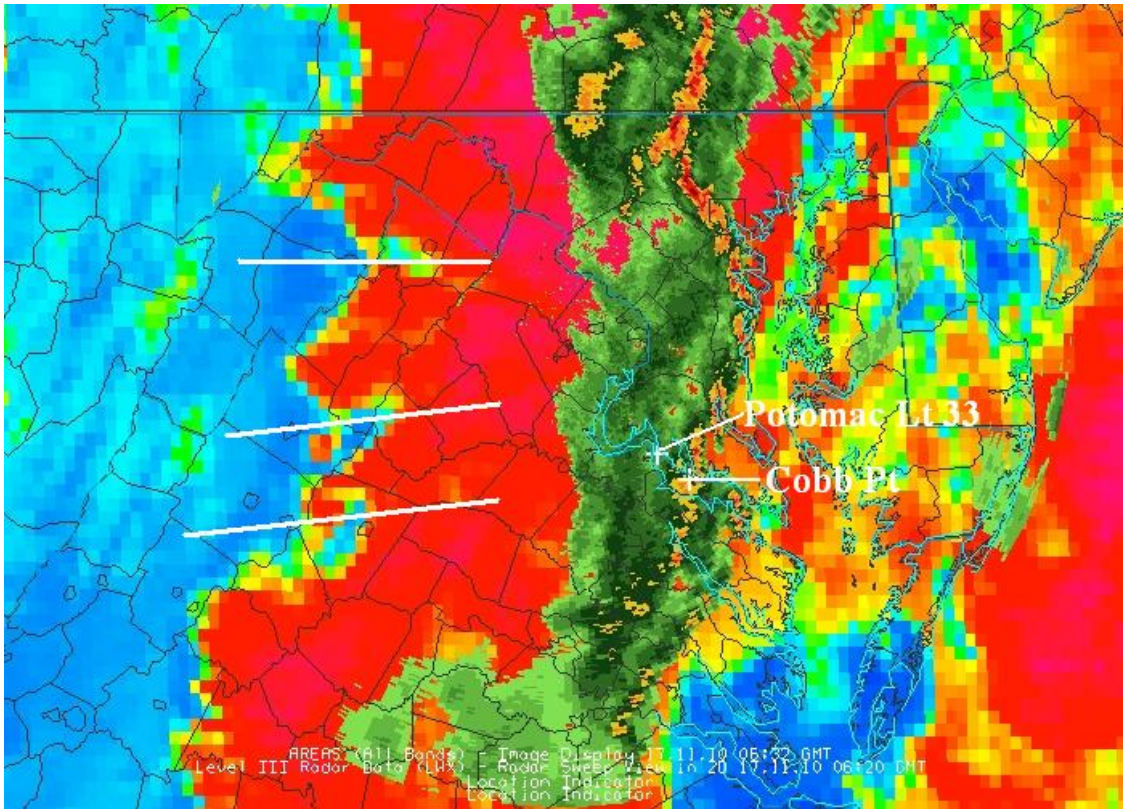

Figure 4. GOES-13 WV-IR BTD image at 0632 UTC 17 November 2010 with overlying radar reflectivity from Sterling, Virginia NEXRAD.

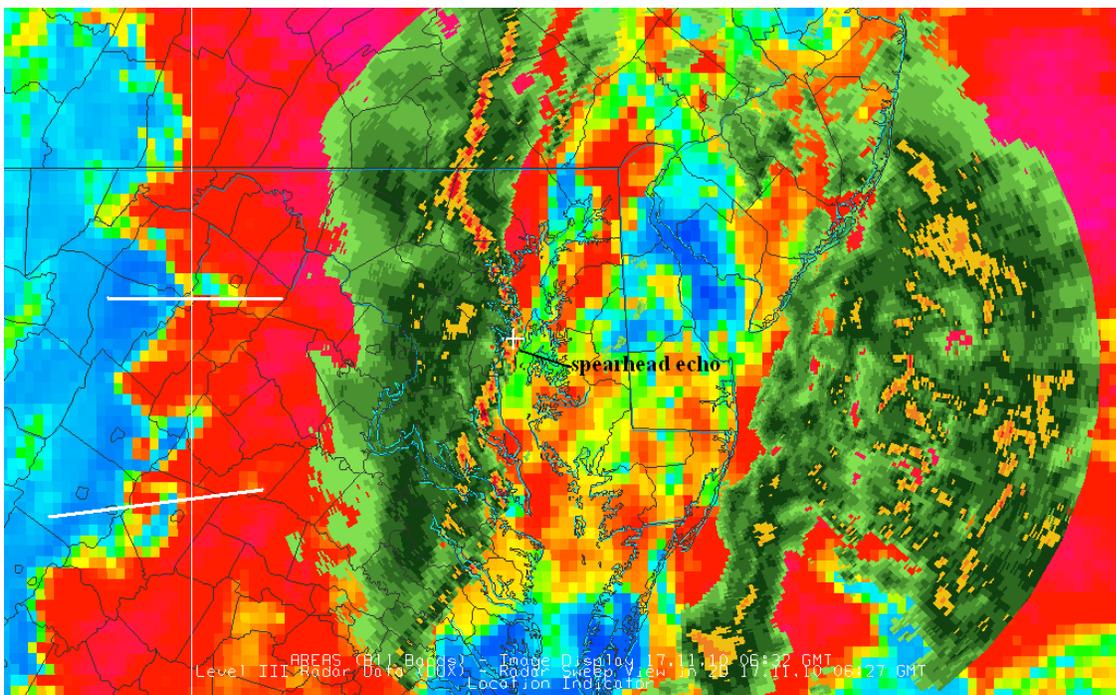

Figure 5. GOES-13 WV-IR BTD image at 0632 UTC 17 November 2010 with overlying radar reflectivity from Dover, Delaware NEXRAD.

Figure 6. Geostationary Operational Environmental Satellite (GOES) imager channel 3 - 4 brightness temperature difference (BTD) product image at 0615 UTC 17 November 2010 with overlying radar reflectivity. "N" and "S" mark the location of Naval Research Laboratory and Silver Spring DCNet observing stations, respectively. White lines mark the location of dry-air notches.

Figure 7. Velocity profile of a microburst from Fujita (1981).

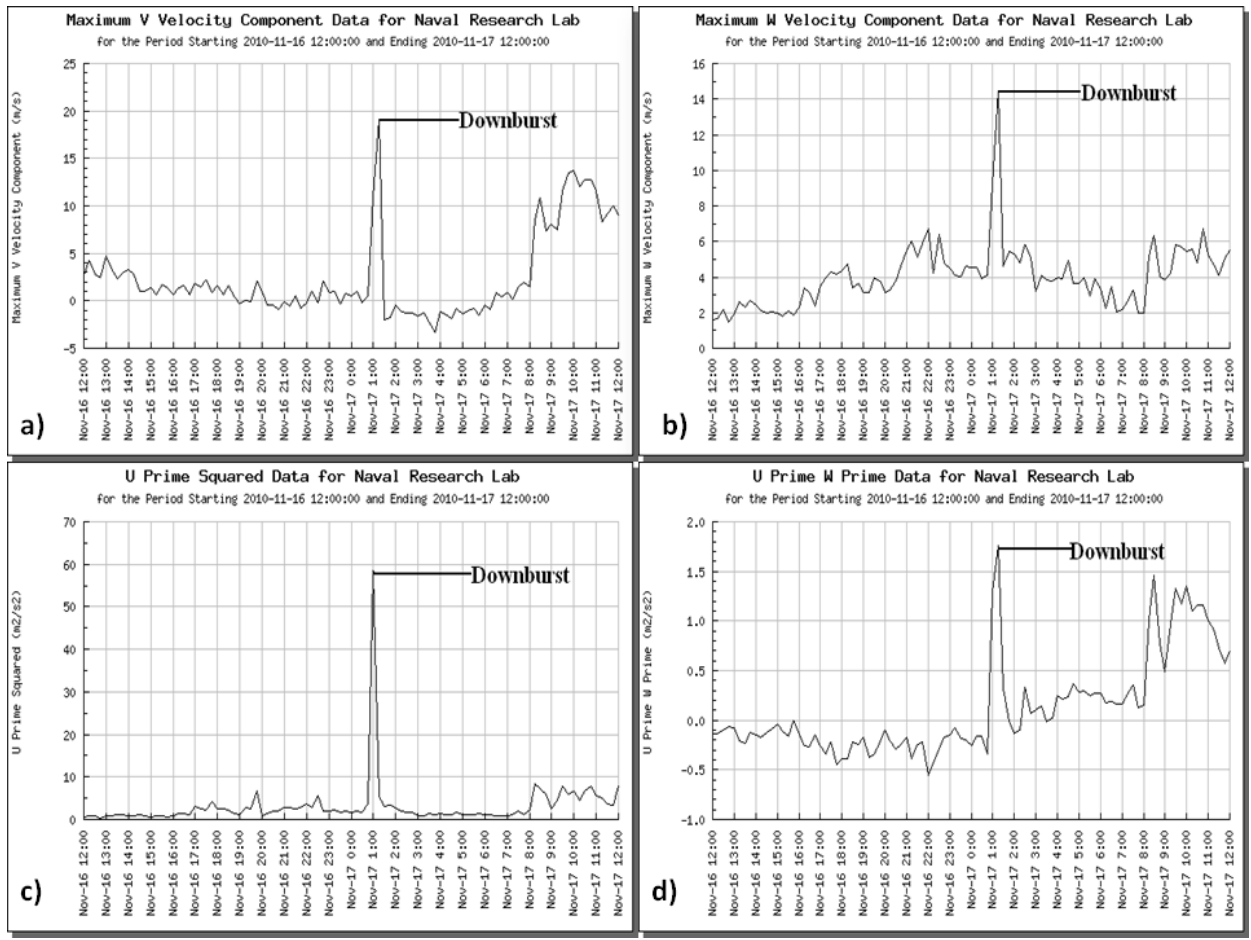

Figure 8. Histograms of a) meridional wind component (V); b) vertical wind component (W); c) wind fluctuation (gustiness); and d) horizontal (u) momentum transport observed at the Naval Research Laboratory DCNet station for 17 November 2010. Note downburst occurrence between 0600 and 0615 UTC (0100 and 0115 EST).

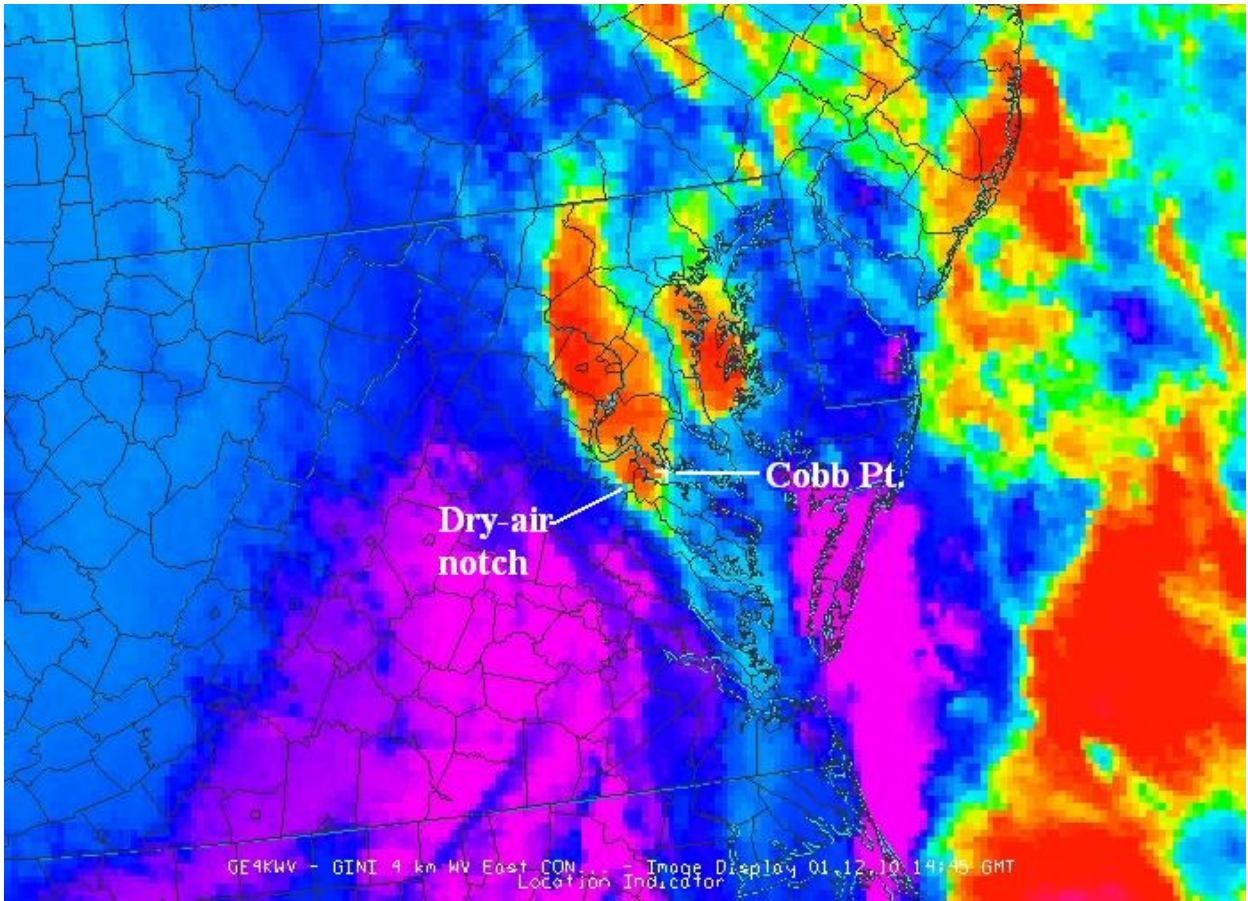
Figure 9. GOES-13 VW-IR brightness temperature difference (BTD) image at 1445 UTC 1 December 2010.

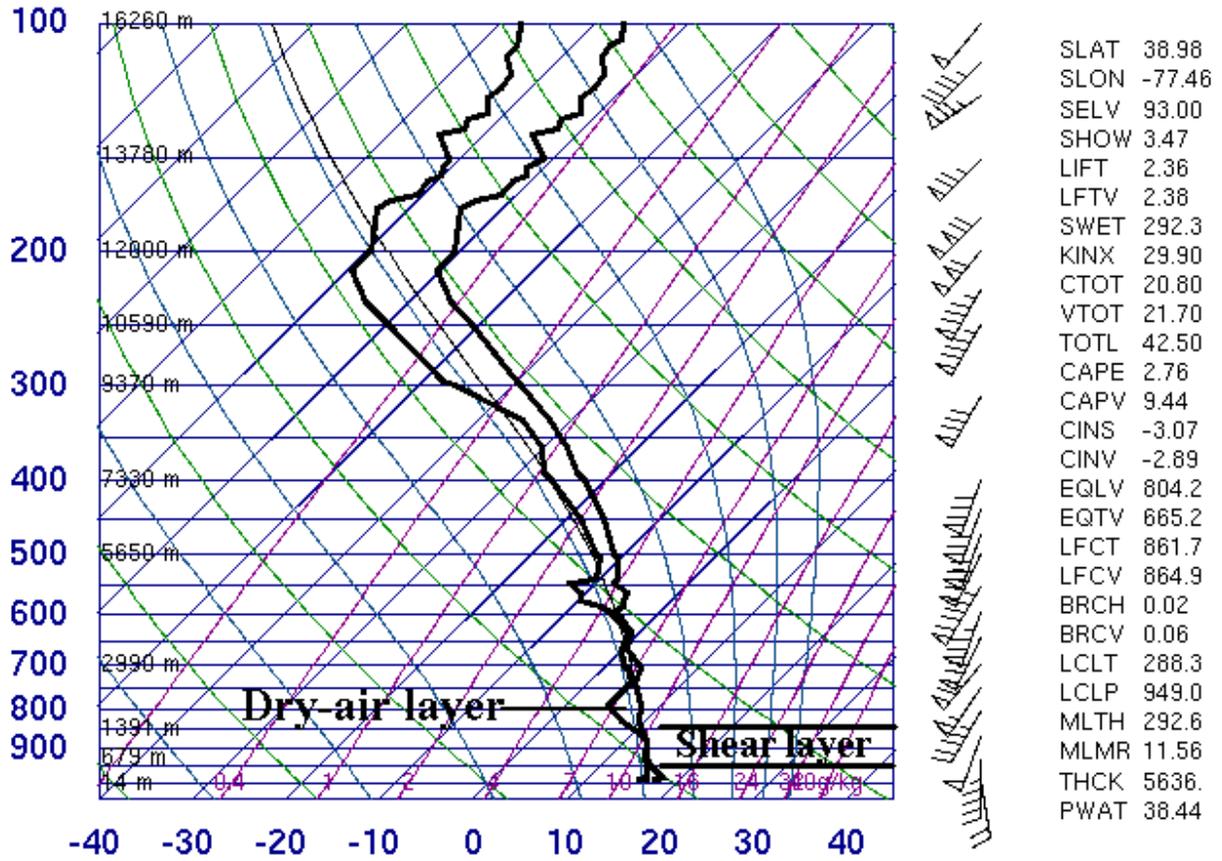
Figure 10. Radiosonde observation (RAOB) from Dulles Airport, VA at 1200 UTC 1 December 2010.

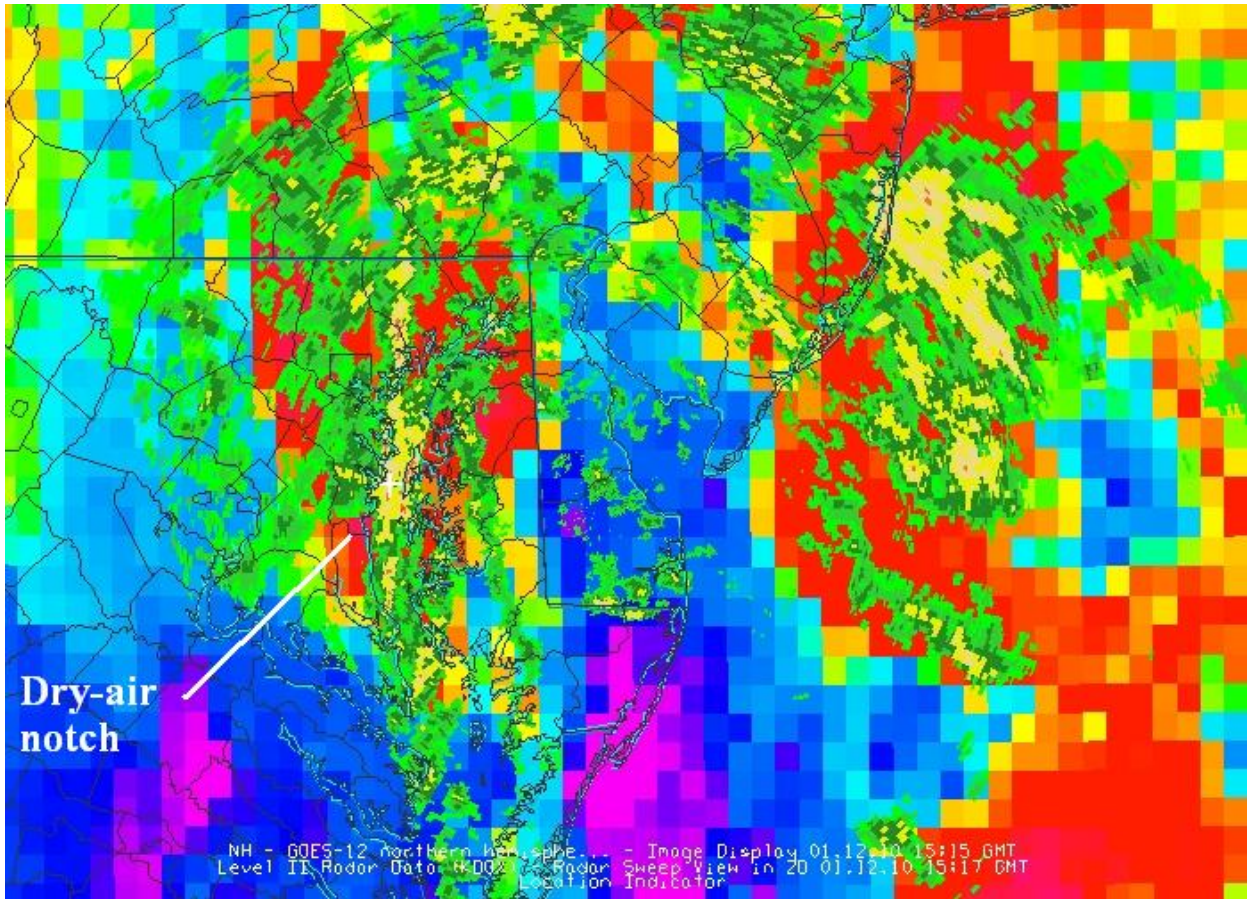

Figure 11. GOES-13 WV-IR BTD image at 1515 UTC 1 December 2010 with overlying radar reflectivity from Dover, Delaware NEXRAD.

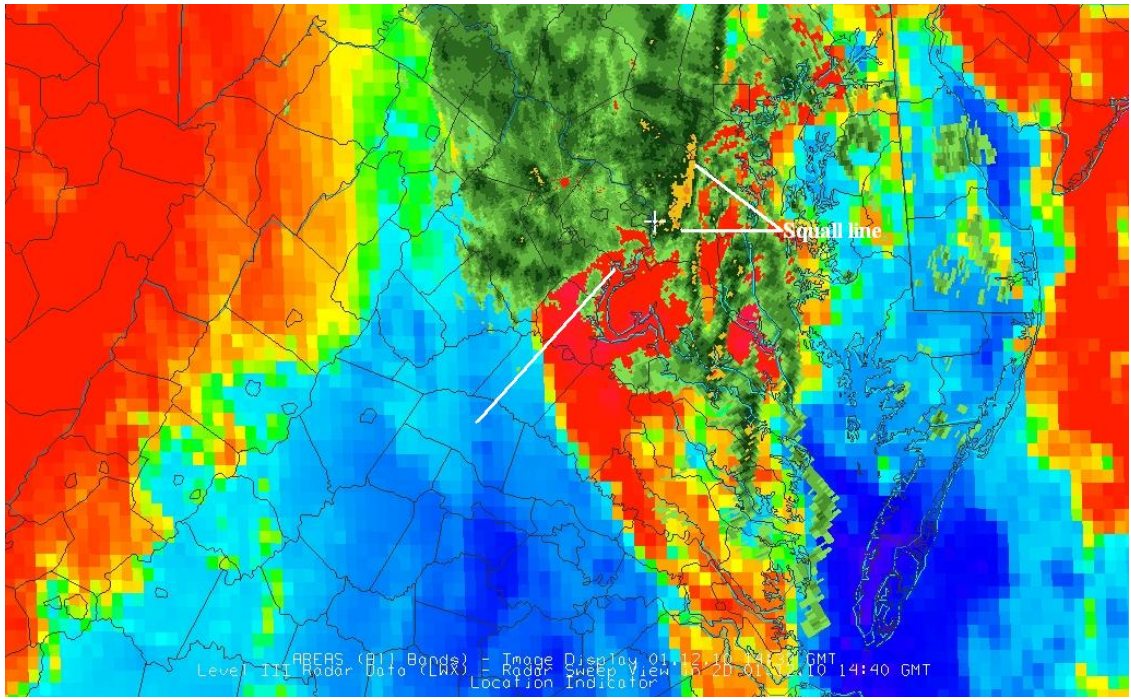

Figure 12. Geostationary Operational Environmental Satellite (GOES) imager channel 3 - 4 brightness temperature difference (BTD) product image at 1432 UTC 1 December 2010 with overlying radar reflectivity. White cross marks the location of Naval Research Laboratory DCNet observing station. White line marks the location of a rear-inflow jet.

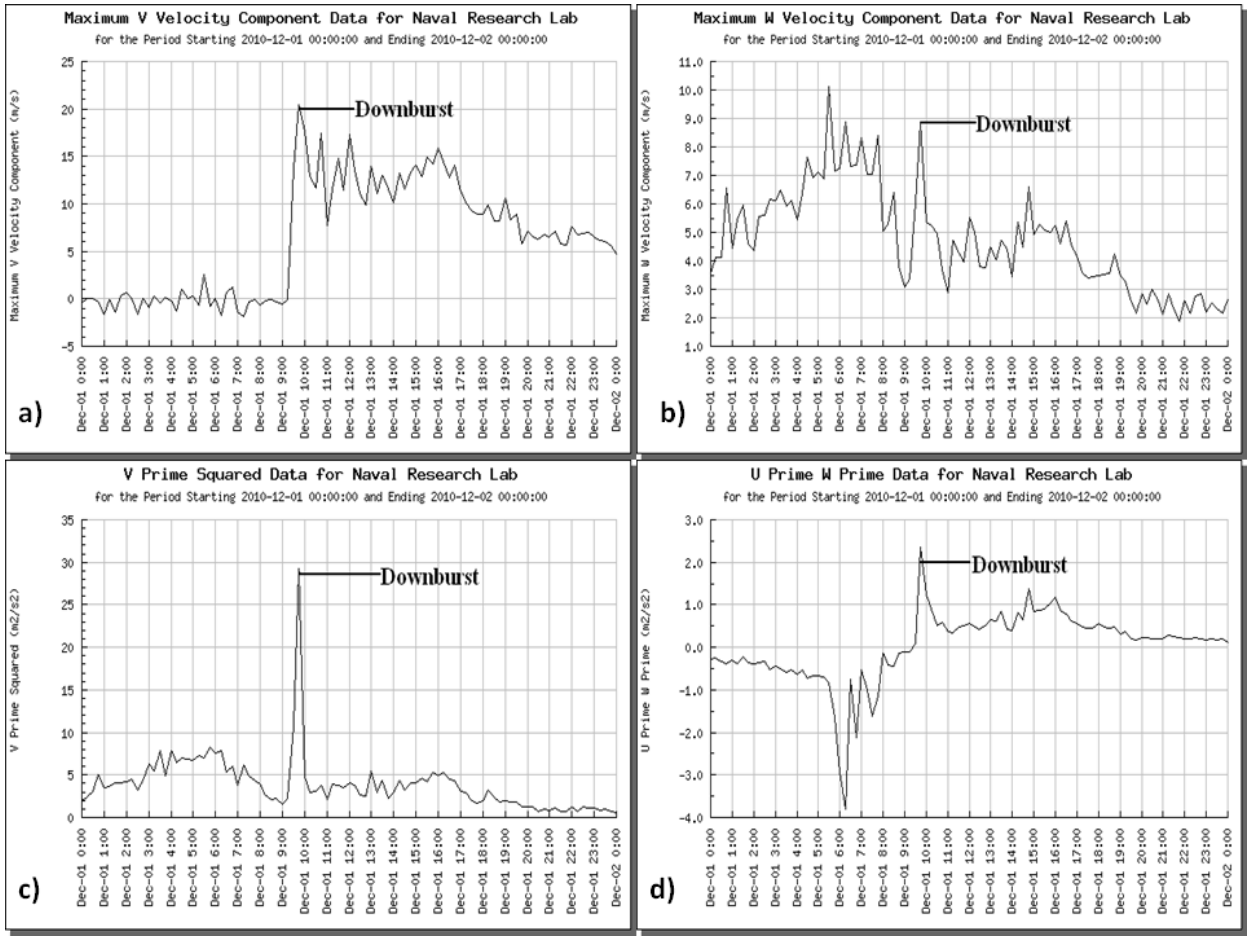

Figure 13. Histograms of a) meridional wind component (V); b) vertical wind component (W); c) wind fluctuation (gustiness); and d) horizontal (u) momentum transport observed at the Naval Research Laboratory DCNet station for 1 December 2010. Note downburst occurrence at 1445 UTC (0945 EST).